\documentclass{aa}
\usepackage{graphics,epsfig,txfonts,color}
\usepackage[dvips]{rotating}
\usepackage{natbib}
\bibpunct{(}{)}{;}{a}{}{,}

\newcommand{\ergcm}[1]{$\times 10^{#1}$ erg cm$^{-2}$ s$^{-1}$}
\newcommand{\ergcmsr}[1]{$\times 10^{#1}$ erg cm$^{-2}$ s$^{-1}$ sr$^{-1}$}

\newcommand{\ergs}[1]{$\times 10^{#1}$ erg s$^{-1}$}

\newcommand{\hcm}[1]{$\times 10^{#1}$ cm$^{-2}$}

\newcommand{\nh}{N$_{\rm H}$}

\newcommand{\fx}{\hbox{F$_{\rm x}$}}
\newcommand{\fxsurf}{\hbox{F$_{\rm x,surf}$}}

\newcommand{\ltsima}{$\buildrel < \over \sim$}
\newcommand{\lsim}{\lower.5ex\hbox{\ltsima}}
\newcommand{\gtsima}{$\buildrel > \over \sim$}
\newcommand{\gsim}{\lower.5ex\hbox{\gtsima}}

\newcommand{\chandra}{\emph{Chandra}}

\newcommand{\terzan}{Terzan~5}

\begin{document}
\definecolor{orange}{cmyk}{0,.5,1,0}
 
\title{Chandra detection of diffuse X-ray emission from the globular cluster \terzan}
\author{P.\,Eger\inst{1}, 
		W.\,Domainko\inst{2}, 
		A.-C.\,Clapson\inst{2}
	   }

\titlerunning{Diffuse X-ray emission from Terzan 5}
\authorrunning{P.\,Eger et al.}

\institute{\inst{1} Erlangen Centre for Astroparticle Physics, 
				Universit\"at Erlangen-N\"urnberg, Erwin-Rommel-Str. 1, 
				91058 Erlangen, Germany\\
		   \inst{2} Max-Planck-Institut f\"ur Kernphysik, PO Box 103980, 
		   		69029 Heidelberg, Germany\\
           \email{peter.eger@physik.uni-erlangen.de}
           }
 
\date{Received / Accepted }
 
\abstract{} 
         {
		 \terzan , a globular cluster (GC) prominent in mass and population of compact
		 objects, is searched for diffuse X-ray emission, as proposed
		 by several models. 
		 }
         {
		 We analyzed the data of an archival \chandra\ observation of \terzan\ to search for extended 
		 diffuse X-ray emission outside the half-mass radius of the GC. 
		 We removed detected point sources from the data to extract spectra from diffuse regions 
		 around \terzan . The Galactic background emission was modeled by a 2-temperature 
		 thermal component, which is typical for Galactic diffuse emission. 
		 }
         {
		 We detected significant diffuse excess emission above the particle background level from the 
		 whole field-of-view. 
		 The surface brightness appears to be peaked at the GC center and decreases smoothly outwards. 
		 After the subtraction of particle and Galactic background, the excess spectrum of the diffuse 
		 emission between the half-mass radius and 3' can be described 
		 by a power-law model with photon index $\Gamma$ = 0.9$\pm$0.5 and a surface flux of 
		 \hbox{\fx\ = (1.17$\pm$0.16)\ergcmsr{-7}} in the 1--7~keV band. 
		 We estimated the contribution from unresolved point sources to the observed excess to be negligible. 
		 The observations suggest that a purely thermal origin of the emission is less likely 
		 than a non-thermal scenario.
		 However, from simple modeling we cannot identify a clearly preferred scenario. 
		 }
         {}

\keywords{globular clusters: individual: Terzan 5 - Radiation mechanisms: non-thermal - Acceleration of particles - X-rays: ISM}
 
\maketitle

\section{Introduction}
\terzan\ \citep[stellar luminosity L$_{\rm star}$=1.5$\times$10$^5$L$_\odot$,][]{1996AJ....112.1487H} is the Galactic 
globular cluster (GC) with the largest population of known millisecond pulsars 
\citep[33 MSPs,][]{2008IAUS..246..291R}. 
Furthermore, a high production rate for low-mass X-ray binaries (LMXBs) is expected in the extremely 
dense core of \terzan\ \citep{2005ASPC..328..231I}. 
The GC is located at a distance of 5.5~kpc \citep{2007A&A...470.1043O} or 8.7~kpc \citep{2002ApJ...571..818C}, 
only $\sim$1.7~deg above the Galactic plane (\hbox{R.A.: 17$^{\rm h}$48$^{\rm m}$04$\fs$0}, \hbox{Dec.: -24$^\circ$46'45"}). 
Its core (r$_{\rm c}$) and half-mass radii (r$_{\rm h}$) are 0.18' and 0.83', respectively 
\citep{1996AJ....112.1487H}. 

GCs are expected to contain intracluster gas originating
from the mass loss of evolved stars. 
Since GCs move through the Galactic halo medium with typical velocities of 
$\sim$200~km~s$^{-1}$, bow shocks should form in front of them in the direction of their proper motion 
\citep{1995ApJ...451..200K}. 
These shocks have the ability to both accelerate particles and heat the gas behind them. 
In this scenario, electrons accelerated at the bow shock could produce diffuse non-thermal X-ray emission 
because of either inverse Compton (IC) scattering on ambient photon fields \citep{1995ApJ...451..200K} 
or non-thermal bremsstrahlung resulting from the deflection of the electrons by inter-stellar medium 
(ISM) nuclei \citep[][henceforth Ok07]{2007PASJ...59..727O}.
Also, diffuse thermal X-ray emission can be emitted from shock-heated material trailing
behind the moving GC. 
For a detailed discussion of the bow-shock scenario, see Ok07. 

The first unresolved diffuse X-ray emission from GCs (47~Tuc, $\omega$~Cen, and M~22) was reported by 
\citet{1982ApJ...254L..11H} using the \emph{Einstein} observatory, which was later 
confirmed by \citet{1995ApJ...451..200K} with \emph{ROSAT}. 
Recently, using \chandra\ data, Ok07 detected significant 
diffuse X-ray emission from the GCs 47~Tuc, NGC~6752, M~5, and $\omega$~Cen. 
They find that the diffuse source at the position of $\omega$~Cen is likely to be a background cluster of galaxies.
A similar extra-Galactic nature is proposed for 47~Tuc by \citet{2009arXiv0906.3583Y}. 
The remaining potentially GC-associated diffuse X-ray emission, 
from M~5 and NGC~6752, could arise from different scenarios. 
The emission in M~5 features an arclike morphology and exhibits a thermal spectrum (kT$<$0.1~keV), possibly from 
shock-heated gas. 
The clumpy structure seen near NGC~6752 presents a hard non-thermal spectrum ($\Gamma\sim$~2) and 
a radio counterpart, maybe from non-thermal bremsstrahlung emission by shock-accelerated electrons 
hitting nearby gas clouds. 

Non-thermal emission in the X-ray band may also be associated with 
compact objects, either directly, as detected from isolated low-mass 
X-ray binary systems \citep[see for instance][for such a candidate system in \terzan]{2005ApJ...618..883W}, 
or as secondary emission from the population of high-energy particles they would generate. 
This was modeled for millisecond pulsars by \citet[][henceforth VJ08]{2008AIPC.1085..277V} 
for the synchrotron radiation mechanism. 
The second case may translate into larger physical scales, due to the diffusion of the energetic particles
away from their source. 
The populations of compact objects in \terzan\ may provide an opportunity to test these scenarios.
Apparent, extended X-ray emission from the direction of GCs might also arise from a population of faint 
unresolved point-like sources below the detection limit of the observing instrument. 

In this work we analyzed an archival \chandra\ observation to search for a diffuse emission component 
above the Galactic background associated with \terzan .
We characterized the X-ray signal using spatially resolved spectral 
analyses, after a careful study of the diffuse Galactic background, and briefly discuss different scenarios 
for the emission. 

\section{X-ray analysis and results}
\subsection{Observation and data preparation}
To search for extended diffuse X-ray emission from \terzan , we analyzed the 
\emph{Advanced CCD Imaging Spectrometer} \citep[ACIS,][]{2003SPIE.4851...28G} data of an 
archival 40~ks \chandra\ \citep{2002PASP..114....1W} observation (ObsID 3798), 
which was originally performed to characterize the faint X-ray point-source population of this GC 
\citep[][henceforth H06]{2006ApJ...651.1098H}. 
Only the ACIS-S3 chip was switched on, so we were only able 
to search for diffuse X-ray emission at angular distances $\lesssim$4' from the cluster core. 
A comprehensive study of diffuse X-ray emission seen from a number of Galactic GCs was performed by 
Ok07. 
However, these authors excluded \terzan\ from their work because the only available \chandra\ dataset at that time 
suffered from serious pile-up effects because of a bright binary outburst in the field-of-view (FoV). 
In this paper we analyzed data from a newer observation where such an event did not occur.

For the X-ray analysis we used the CIAO software version 4.1, supported by tools from the FTOOLS package 
and XSPEC version 12.5.0 for spectral modeling \citep{1996ASPC..101...17A}. 
The \emph{event1} data were reprocessed with the latest position and energy calibration (CTI correction, v4.1.3) 
using bad pixel files generated by \emph{acis\_run\_hotpix}. 
The good-time-interval (GTI) file supplied by the standard processing, which was used by H06, 
screens out a $\sim$4.0~ks interval of strong background flaring at the end of the observation. 
To remove an additional time period of 4.3~ks with a slightly increased background level, we 
used the light curve in the 0.5--7.0~keV energy band after the core region of the cluster and additional 
bright sources were removed from the data. 
A screening threshold of 1.0~cts/s yielded a net exposure of 31.0~ks. 
We chose these stricter criteria with respect to H06 because understanding the background 
is crucial for analyzing faint extended sources. 

\subsection{Extraction regions}
To detect and remove point-like X-ray sources from the event-list, we ran \emph{wavdetect} on the GTI-screened dataset 
in three energy bands (0.5--2.0~keV, 2.0--7.0~keV and 0.5--7.0~keV). 
H06 used \emph{pwdetect} for the detection within r$_{\rm h}$ and \emph{wavdetect} for outer regions. 
In this paper we only analyzed areas outside r$_{\rm h}$, so results should be comparable. 
We estimated a point-source detection limit of $\sim$2\ergcm{-15} in the 0.5--7.0~keV band. 
For the most part our results are compatible with the sources listed by H06, Table~2. 
However, the shorter exposure time compared to the analysis of H06 led to a higher 
point-source detection threshold. 
Therefore we did not detect the faintest seven sources from H06 that we introduced manually into 
our source list. 
Sources were removed from the dataset using the 3$\sigma$ radius of the point spread function. 
Additionally, all events within r$_{\rm h}$ were disregarded. 

\begin{figure}
  \resizebox{0.98\hsize}{!}{\includegraphics[clip=]{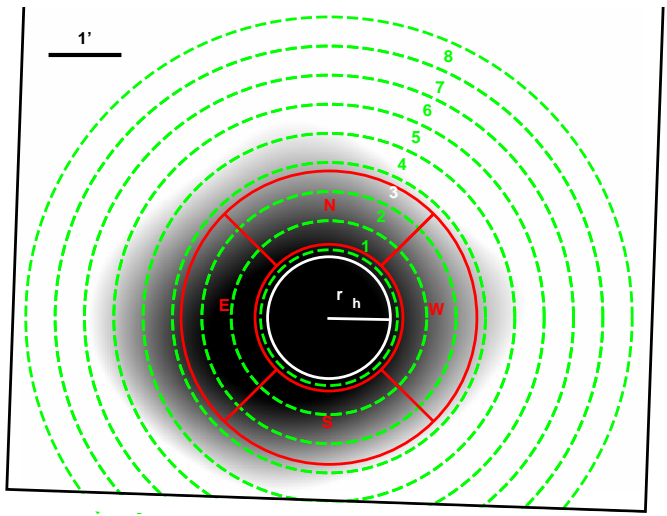}}
  \caption{
  		Smoothed, exposure corrected and NXB subtracted \chandra\ image of diffuse X-ray emission in the 
		1--7~keV band around \terzan . 
  		Excluded regions from point sources and the region inside r$_{\rm h}$ (white circle labeled r$_{\rm h}$) 
		were refilled (see text). 
		The wings seen towards the north, east and west are only marginally significant and might 
		be artifacts of the smoothing algorithm. 
		The FoV for this observation is drawn as a box (black). 
		Shown are the eight annular extraction regions (dashed green lines, numbered 1 to 8) and the 
		four pie-shaped regions (solid red lines, labeled N, E, S and W). 
		The color scaling is linear and chosen such that the Galactic diffuse level is saturated as white. 
		Thus only emission above the Galactic diffuse level appears in gray scales. 
		}
  \label{fig-map}
\end{figure}

To measure the level of diffuse X-ray emission around 
\terzan , we extracted spectra from eight concentric annular regions centered on the cluster core 
with radii from 1.1' to 3.9' (Fig.~\ref{fig-map}). 
Each ring has a width of 0.4'. 
We chose rings with equal width over rings with constant area to have comparable statistical quality 
in the spectra since the surface brightness decreases with distance from the GC.  
For the spectral analysis, we chose the 1--7~keV energy band. 
Widening the band in either direction lead to lower signal-to-noise ratios. 
At lower energies an increased contribution to the signal from soft thermal Galactic diffuse 
emission is expected. 
At energies above 7--8~keV the charged particle induced background component 
increases significantly for instruments onboard \chandra . 
The mean effective area and energy response for each spectrum was calculated by weighting the contribution 
from each pixel by its flux using a detector map in the same energy band. 
To subtract the particle induced non-X-ray background (NXB), we used a background dataset provided 
by the calibration database, where the detector was operated in stowed position. 
The background spectrum for each ring was extracted from the respective region in this background 
dataset. 
To account for the time dependence of the NXB, we scaled the background by the ratio 
of the source and background count rates in the 9--12~keV energy band for each spectrum 
\citep[as described by][]{2003ApJ...583...70M}. 

To produce an image of diffuse X-ray emission above the particle background from the direction of \terzan , 
we extracted counts in the 1--7~keV energy band and refilled the excluded source regions and the region 
inside r$_{\rm h}$ with \emph{dmfilth} using the photon distributions from rings around the excluded areas. 
We subtracted the particle background using the respective image from the stowed dataset after correction 
for the different exposures. 
The resulting image was corrected for relative exposure and adaptively smoothed with \emph{asmooth}. 
We required a minimum significance of 3$\sigma$ for the kernel size of the smoothing algorithm. 
The resulting smoothing radii were a few arcminutes, so that no details smaller than that 
scale can be seen in the smoothed image. 
Figure~\ref{fig-map} shows the resulting image together with all extraction regions that we used in this work.

\begin{figure}
  \resizebox{0.98\hsize}{!}{\begin{turn}{-90}\includegraphics[clip=]{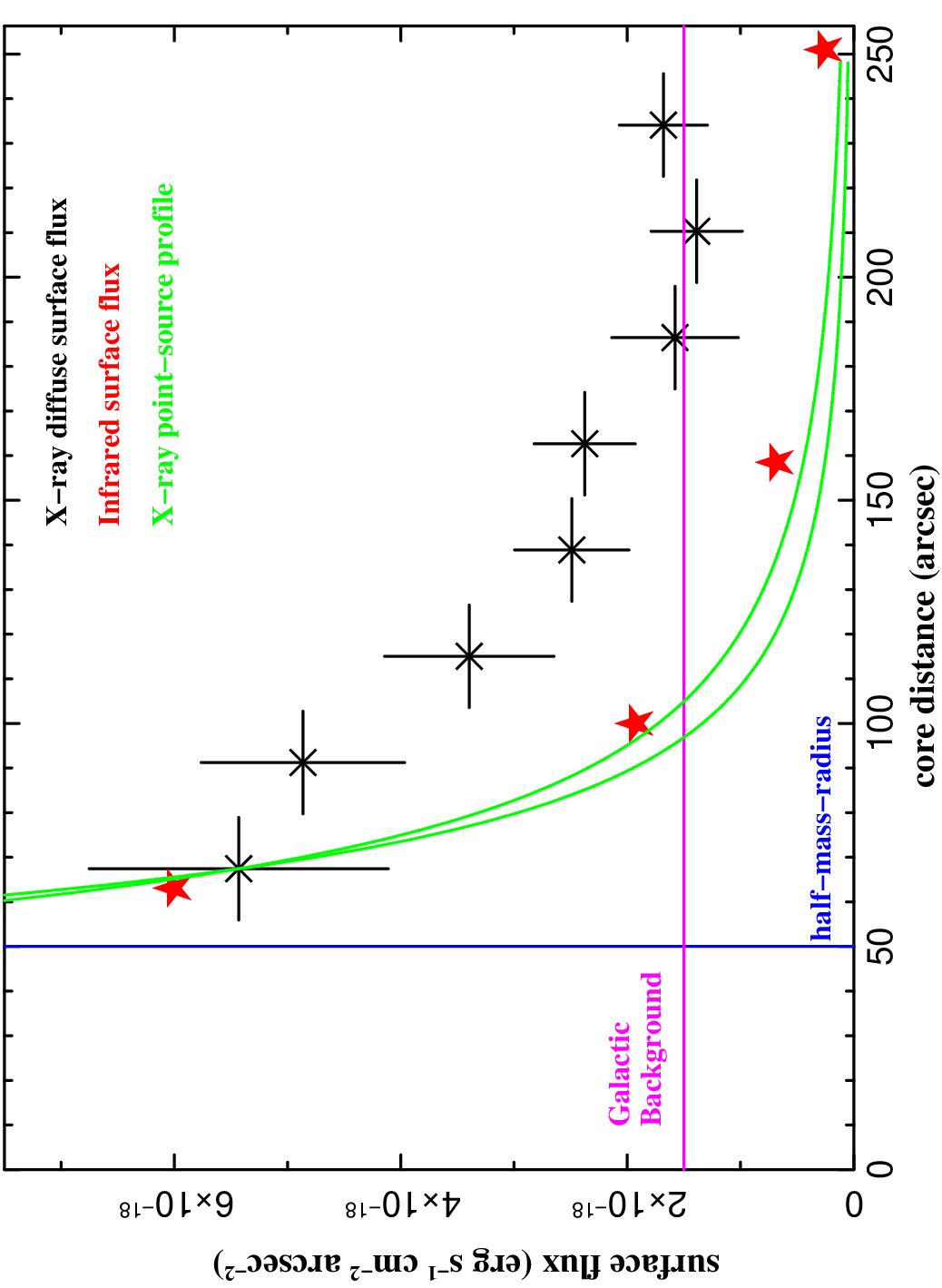}\end{turn}}
  \caption{Radial dependence of the observed diffuse X-ray surface flux above the particle background 
  		in the 1--7\,keV band as seen with \chandra\ (black crosses with error bars). 
		Stars (red) denote the infrared surface brightness profile from \citet{1995AJ....109..218T}. 
		The solid curves (green) show the X-ray point-source distribution described by 
		a generalized King-profile \citep{2006ApJ...651.1098H} for the two extreme cases 
		q=1.43$\pm$0.11. 
		All profiles are scaled to match the first diffuse X-ray data point, using an exponential fit 
		in the case of the infrared data. 
		The vertical (blue) and the horizontal (magenta) line denote r$_{\rm h}$ 
		\citep{1996AJ....112.1487H} and the Galactic diffuse background level, respectively.}
  \label{fig-radial-flux}
\end{figure}

Even though a significant contribution from thermal Galactic diffuse emission is expected, 
the spectra from the single rings were fit well enough by an absorbed power-law model 
for a preliminary flux estimate. 
The resulting fit parameters are given in Table~\ref{tab-specparams}. 
We found significant diffuse excess emission above the particle background in all rings and 
derived the surface brightness for each region by dividing the model flux by the effective extraction 
area, which is the geometric ring area inside the FoV minus excluded regions and bad pixels.  
Due to limited statistics, we fixed the column density at a default value of \hbox{\nh\ = 1\hcm{22}}. 
Therefore, we list the observed surface fluxes in Table~\ref{tab-specparams}, as opposed to the intrinsic fluxes 
we provide for all other spectra. 
The diffuse surface flux shows a clear radial dependence (Fig.~\ref{fig-radial-flux}), 
which indicates that a significant part of the excess is connected to the cluster. 
At distances greater than $\sim$170" from the GC core, the observed surface flux seems to reach a base level of 
\fxsurf\ $\approx$ 1.5\ergcm{-18}arcsec$^{-2}$ (1--7~keV). 
In the following section we derive the unabsorbed surface flux for the outer region by applying 
a more realistic physical model.

\subsection{Galactic diffuse background}
\terzan\ is close to the Galactic plane where diffuse Galactic emission becomes an important component. 
However, the \chandra\ blank-sky datasets are composed of observations towards high Galactic latitudes, 
which would underestimate the sky background in our case. 
To test whether the spectrum observed from the outer three rings is compatible with 
thermal Galactic diffuse emission, we used a more physically reasonable model. 
Similar to \citet{1997ApJ...491..638K} and \citet{2005ApJ...635..214E}, who modeled the diffuse 
Galactic ridge emission as observed with \emph{ASCA} and \chandra , respectively, 
we describe the Galactic diffuse component using a two-temperature (2-T) non-equilibrium 
ionization model (NEI) \citep{1984Ap&SS..98..367M}. 
To improve the statistical quality, we combined the outer three (175--246") rings into a single spectrum. 
The spectrum was adaptively binned to a minimum of 20 excess counts per bin. 
As background we again used the spectrum extracted from the same region in the NXB dataset. 
We fitted a 2-T NEI model to the outer spectrum, freezing most of the parameters to the 
best-fit values from Table~8 in \citet{2005ApJ...635..214E}. 
We left the surface brightnesses of the two components and the \nh\ free to vary to account for the difference 
in flux and column density between the region around \terzan\ and the area observed by \citet{2005ApJ...635..214E}. 
In addition, we allowed the Si-abundance of the soft component as a free fit-parameter, because the 
low-ionized Si line at $\sim$1.8~keV \citep{1997ApJ...491..638K,2005ApJ...635..214E} was otherwise underestimated. 

The spectrum of the outer region together with the model fit is shown in Fig.~\ref{fig-spectra} (\emph{Top}). 
To be able to compare our results to the analysis of \citet{2005ApJ...635..214E}, we chose an energy range of 0.7--10~keV 
in this specific case. 
The best-fit values are given in Table~\ref{tab-specparams} (\emph{Outer} region). 
The total intrinsic surface flux of the two components is a factor of three lower than the value for the Galactic region 
observed by \citet{2005ApJ...635..214E}. 
This relation is in good agreement with the ratio between the column densities for both regions, which 
is $\sim$4 \citep{1990ARAA...28..215D}. 
Assuming that the Galactic column density seen from a certain direction is directly related to the 
expected flux from a diffuse Galactic component, we conclude that at least $\sim$3/4 of the total excess above particle 
background observed from the outer region comes from Galactic diffuse emission. 

\begin{table*}
\caption[]{Extraction regions and results from spectral fitting}
\renewcommand{\tabcolsep}{3.5pt}
\begin{center}
\begin{tabular}{lllllllll}
\hline\hline\noalign{\smallskip}
\multicolumn{1}{l}{Region} &
\multicolumn{1}{l}{Distance range$^{(1)}$} &
\multicolumn{1}{l}{Angular range$^{(2)}$} &
\multicolumn{1}{l}{Excess counts$^{(3)}$} &
\multicolumn{1}{l}{NEI: \fxsurf$^{(4)}$} &
\multicolumn{1}{l}{NEI: kT$^{(5)}$} &
\multicolumn{1}{l}{PL: \fxsurf$^{(6)}$} &
\multicolumn{1}{l}{PL: $\Gamma ^{(7)}$} &
\multicolumn{1}{l}{$\chi^{\scriptscriptstyle{2}}_{\scriptscriptstyle{\nu}}$(d.o.f.)} \\
\multicolumn{1}{l}{} &
\multicolumn{1}{l}{(arcsec)} &
\multicolumn{1}{l}{(deg)} &
\multicolumn{1}{l}{} &
\multicolumn{1}{l}{\tiny ($10^{-7}$ erg cm$^{-2}$ s$^{-1}$ sr$^{-1}$)} &
\multicolumn{1}{l}{(keV)} &
\multicolumn{1}{l}{\tiny ($10^{-7}$ erg cm$^{-2}$ s$^{-1}$ sr$^{-1}$)} &
\multicolumn{1}{l}{} &
\multicolumn{1}{l}{} \\
\noalign{\smallskip}\hline\noalign{\smallskip}
Ring 1	& 55--79	& 0--360	& 195.2$\pm$26.5	& -- & -- & 1.01$\pm$0.25	& 1.8$\pm$0.4 & --/1.3(15) \\
Ring 2	& 79--103	& 0--360	& 278.6$\pm$31.1	& -- & -- & 0.91$\pm$0.17	& 2.1$\pm$0.5 & --/0.7(26) \\
Ring 3	& 103--126	& 0--360	& 274.8$\pm$33.3	& -- & -- & 0.64$\pm$0.14	& 2.2$\pm$0.6 & --/1.2(27) \\
Ring 4	& 126--150	& 0--360	& 259.9$\pm$34.2	& -- & -- & 0.47$\pm$0.09	& 3.0$\pm$0.8 & --/0.9(24) \\
Ring 5	& 150--174	& 0--360	& 276.7$\pm$35.3	& -- & -- & 0.44$\pm$0.08 	& 2.9$\pm$0.7 & --/1.4(26) \\
Ring 6	& 174--198	& 0--360	& 192.7$\pm$34.7	& -- & -- & 0.30$\pm$0.10	& 2.8$\pm$0.8 & --/0.9(19) \\
Ring 7	& 198--222	& 0--360	& 227.5$\pm$36.2	& -- & -- & 0.26$\pm$0.08 	& 3.6$\pm$1.2 & --/1.2(22) \\
Ring 8	& 222--246	& 0--360	& 240.3$\pm$37.5	& -- & -- & 0.31$\pm$0.07 	& 3.5$\pm$1.2 & --/1.0(23) \\
\noalign{\smallskip}\hline\noalign{\smallskip}
Inner	& 55--174	& 0--360 	& 1273.5$\pm$115.8	& 1.34$\pm$0.14 (soft) &  0.59 (soft)   & 1.17$\pm$0.16 & 0.9$\pm$0.5  	& 1.1(52)/ \\
		&			& 0--360 	& 					& 5.36$\pm$0.80 (hard) &  5.0 (hard)    & --  			& --			& 1.3(50) 	\\
Outer	& 175--246	& 0--360 	& 825.3$\pm$62.5	& 0.58$\pm$0.19 (soft) &  0.59 (soft)   & --  			& --			& 1.2(36)/-- \\
		&			& 0--360 	& 					& 2.1$\pm$0.7 (hard)   &  5.0 (hard)    & --  			& --			&   	  \\
\noalign{\smallskip}\hline\noalign{\smallskip}
North  	& 60--120	& 45--135	& 191.0$\pm$26.4	& -- 					& --  			& 1.44$\pm$0.63  & 1.5$\pm$1.0 &  --/1.0(45)  \\
East   	& 60--120	& 135--225  & 176.1$\pm$26.7	& -- 					& --  			& 2.10$\pm$0.70  & 0.6$\pm$0.7 &  --/0.9(44)  \\
South  	& 60--120	& 225--315  & 230.1$\pm$27.0	& -- 					& --  			& 2.46$\pm$0.72  & 0.8$\pm$0.7 &  --/0.8(47)  \\
West   	& 60--120	& 315--45	& 165.2$\pm$27.1	& -- 					& --  			& 2.10$\pm$0.75  & 1.7$\pm$1.1 &  --/1.0(45)  \\
\hline\noalign{\smallskip}
\end{tabular}
\label{tab-specparams}
\end{center}
$^{(1)}$Inner and outer radii of the region. 
$^{(2)}$Angular range of the region. 
$^{(3)}$Excess counts after background subtraction. 
$^{(4)}$Intrinsic 0.7--10.0~keV surface flux of the two thermal components. 
$^{(5)}$Temperatures of the two thermal components. 
$^{(6)}$Surface flux (1--7~keV) resulting from an absorbed power-law fit. 
$^{(7)}$Spectral index resulting from an absorbed power-law fit. 
\end{table*}

\subsection{Diffuse excess emission connected to \terzan }
In this section we focus on the diffuse emission observed from the inner five rings 
(55--175"). 
In addition to the radial dependence of the diffuse excess emission, Fig.~\ref{fig-radial-flux} 
shows the infrared surface brightness profile \citep{1995AJ....109..218T} 
and the X-ray point-source distribution \citep[King-profile from ][]{2006ApJ...651.1098H}. 
Both profiles are scaled to match the first diffuse X-ray data point, using an exponential fit 
in the case of the infrared data. 

\begin{figure}
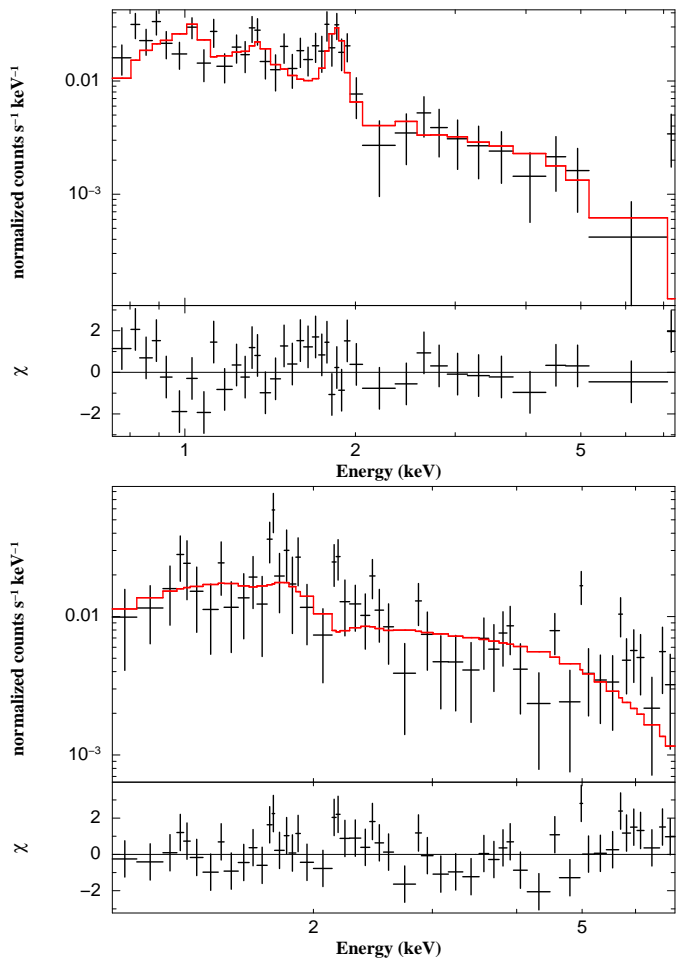

  \resizebox{0.98\hsize}{!}{\begin{turn}{-90}\includegraphics[clip=]{spectrum_outer_8-10.ps}\end{turn}}
  \resizebox{0.98\hsize}{!}{\begin{turn}{-90}\includegraphics[clip=]{spectrum_inner-outer.ps}\end{turn}}
  \caption{
  		\emph{Top}: \chandra\ spectrum from the outer annulus (175--246") with a 2-temperature 
		non-equilibrium ionization model fit (stepped red line). 
		All parameters are fixed to the values from \citet{2005ApJ...635..214E} 
		except for the surface brightnesses of the two components. 
		\emph{Bottom}: \chandra\ spectrum from the inner annulus (55--175") with the spectrum from the 
		outer annulus subtracted as background. 
		The fit is an absorbed power-law model (red stepped line). 
		The parameters for both fits are given in Table~\ref{tab-specparams}.
		}
  \label{fig-spectra}
\end{figure}

To investigate the nature of the diffuse excess emission observed from the inner region in more detail 
and to improve the statistical quality, we extracted the combined spectrum from the inner five rings (55--175"). 
As a first step we fitted the same 2-T NEI model to the NXB subtracted inner spectrum, 
binned to a minimum of 20 excess counts per bin, as was done for the outer region in the previous section. 
The resulting surface fluxes of the two components are listed in Table~\ref{tab-specparams} (\emph{Inner} region). 
Following the same argument as in the previous section, we estimate that in this case only $\sim$1/3 of the 
total observed emission is of diffuse Galactic origin. 
Together with the surface brightness showing a clear radial dependence 
with respect to the core of \terzan , we conclude that a significant part of the observed flux is connected 
to the GC. 

As an estimate for the Galactic diffuse background component, we subtracted the outer 
(175--246", see previous section) from the inner spectrum. 
Figure~\ref{fig-spectra} (\emph{Bottom}) shows the 
resulting excess spectrum from the inner region, binned to a minimum of 20 excess counts per bin, 
together with an absorbed power-law model fit. 
The spectral parameters are collected in Table~\ref{tab-specparams} (\emph{Inner} region). 
Fitting a thermal plasma (MEKAL, $\chi^{\scriptscriptstyle{2}}_{\scriptscriptstyle{\nu}}$ = 1.4(50)) 
or a thermal bremsstrahlung model (BREMSS, $\chi^{\scriptscriptstyle{2}}_{\scriptscriptstyle{\nu}}$ = 1.4(50)) to the spectrum 
gave temperatures kT$>$17~keV and kT$>$25~keV, respectively. 
The most prominent feature of the Galactic thermal emission 
observed from the outer region is an emission line, centered on $\sim$1.8~keV. 
Introducing a Gaussian line at that energy or an additional thermal component to the 
excess spectrum from the inner annulus does not significantly improve the fit ($\chi^{\scriptscriptstyle{2}}_{\scriptscriptstyle{\nu}}$ = 1.2). 
Therefore, we conclude that the spectrum from the outer region describes the Galactic diffuse background component 
sufficiently. 
The total unabsorbed diffuse excess flux in the 1--7~keV band measured from the inner region above the Galactic 
background is F$_{\rm X}$~=~(5.5$\pm$0.8)\ergcm{-13}. 
Assuming a distance of 5.5~kpc the intrinsic luminosity is L$_{\rm X}$~=~(2.0$\pm$0.3)\ergs{33}. 

We found no indication of a variation in the spectral index with increasing radius, when 
subdividing the inner region into two or more sub-regions. 
Furthermore, there is no evidence of a directional variation in the index and flux. 
Table~\ref{tab-specparams} (\emph{north, east, south, and west} regions) 
illustrates the result for directional dependence of spectra extracted from pie-shaped regions 
towards the north, east, south and west with respect to the cluster center (Fig.~\ref{fig-map}). 
Their inner and outer radii are 60" and 120", respectively. 
The latter value was chosen such that the southern region is not truncated by the FoV. 
As background we again used the spectrum from the outer annulus. 
A similar result was achieved when the regions were rotated by 45~degrees. 

\section{Origin of the diffuse emission}
The present results indicate GC-centered diffuse hard X-ray excess emission 
above Galactic background, which extends significantly beyond r$_{\rm h}$. 
In this section we briefly discuss standard thermal and non-thermal emission scenarios, leaving 
out more exotic possibilities, as described by, e.g., \citet{2005A&A...444L..33D}. 
Throughout this section we use a distance to \terzan\ of 5.5~kpc. 
The larger distance estimate (8.7~kpc) would increase the energy requirements for the models by a factor of 2.5. 

\subsection{Contribution from unresolved point sources}
The luminosity of unresolved point sources inside r$_{\rm h}$ has been estimated by 
\citet{2006ApJ...651.1098H} to 8\ergs{32}. 
They furthermore determined the spatial surface distribution of X-ray 
sources in \terzan\ to $S(r)\propto(1 + (r/r_c)^2)^{(1-3q)/2}$ with q=1.43. 
From this distribution we expect to find in the 1--3 arcmin annulus 
only 9\% of the luminosity of unresolved point sources within r$_{\rm h}$. 
The expected 7\ergs{31} is much lower than the measured emission 
(2.0$\pm$0.3)\ergs{33}, so we conclude that the contribution from unresolved point sources is negligible. 

\subsection{Synchrotron radiation}
One possibility for producing non-thermal emission by relativistic electrons is synchrotron radiation emission (SR), 
which would radiate at a frequency 
$\nu_{\rm syn} $=$ 120 \, (\gamma/10^4)^2 \, (\mathrm{B}/1 {\rm \mu G})$ sin$(\phi)$~MHz, 
where B is the strength of the magnetic field, $\gamma$ the Lorentz factor of the electrons, and $\phi$ 
the pitch angle between the magnetic field and the electron velocity (Ok07). 
Following Ok07, electrons with an energy of $\sim$10$^{14}$~eV would be needed to produce SR emission in typical 
Galactic magnetic fields of a few $\mu$G in the keV regime. 
The population of MSPs in the center of \terzan\ 
was suggested as a continuous source of such highly-energetic electrons \citep{2007MNRAS.377..920B,2009ApJ...696L..52V}. 
These particles propagate to the observed extension of the diffuse emission of 3' (4.8~pc) on a timescale 
of t$_{\rm diff}$=$3\times 10^3 \, \mathrm{B}_{\rm 1\mu G}$ years, assuming Bohm diffusion (VJ08). 
The cooling of electrons with energies 
of $\sim10^{14}$~eV in GCs is dominated by SR emission with typical cooling times of 
t$_{\rm cool}\approx$3$\times 10^{4}\, \mathrm{B}_{\rm 1\mu G}^{-2}$~years (VJ08). 
Assuming an injection spectrum with index -1, SR cooling, which depends linearly on the energy of 
the electrons, should change the index to -2. 
Since no such steepening of the spectrum is observed at the 2$\sigma$ level, t$_{\rm diff}$ $\lesssim$ t$_{\rm cool}$ 
is required, which would limit the magnetic field to $\sim$1 $\mu$G or it 
would indicate a faster diffusion of electrons. 
In this scenario, the population of highly energetic electrons has 
to radiate the observed X-ray luminosity (2\ergs{33}) on a timescale of 
t$_{\rm cool}$, so would require a total energy in these electrons of 
1.8~$\times$ 10$^{45}\,\mathrm{B}_{\rm 1\mu G}^{-2}$~erg. 
Associated IC radiation in the TeV energy range should be detectable in spatial coincidence in the 
case of low magnetic fields $\mathrm{B} \lesssim$ few ${\rm \mu}$G, providing a test for this 
scenario (VJ08). 

\subsection{Inverse Compton emission}
Non-thermal X-ray emission in GCs can also be produced by IC up-scattering of star-light 
photons by mildly relativistic electrons \citep{1995ApJ...451..200K}. 
The bow shock of the GC could provide these electrons (Ok07). 
The power of IC radiation $P_{\rm IC}$ emitted by a single electron is given by 
$P_{\rm IC} = 4/3 \,\sigma_T \,c \,\gamma^2 \,u_{\rm rad}$, where $\sigma_T$ is the Thomson cross section, 
$c$ is the speed of light, $\gamma$ the energy of the electrons, and $u_{\rm ph}$ the density 
of the target photon field \citep{1995ApJ...451..200K}. 
Therefore the intensity of IC emission is directly related to the energy density of the target photon field. 
GCs exhibit a very high stellar density in their core region, which decreases rapidly in their outskirts, 
resulting in a centrally peaked photon field as indicated with the distribution of the infrared surface
brightness in Fig.~\ref{fig-radial-flux}. 
The diffuse X-ray emission presented in this paper exhibits a 
surface brightness profile that is roughly similar to this proxy of the density of the photon field, 
consistent with IC emission. 
The energy density $u_{\rm rad}$ of the stellar photon field scaling with 
$u_{\rm rad}\approx L_{\rm star}$/(4$\pi$r$^2$c) 
is about 40~eV/cm$^3$ and 5~eV/cm$^3$, at distances of 1' and 3', respectively, from the center of the GC. 
For the quantitative estimate 
of the total energy in electrons we adopt the model of \citet{1995ApJ...451..200K} giving 
5~$\times$~10$^{49}$~($u_{\rm rad}$/~40~eV~cm$^{-3}$)~erg. 
In this scenario the X-ray emission should be accompanied by potentially detectable SR 
emission in the radio band.

\subsection{Non-thermal bremsstrahlung}
One additional emission process of non-thermal X-rays is non-thermal bremsstrahlung, which is produced when energetic 
electrons are deflected by protons and nuclei. In this scenario the flux of the emission should follow the distribution 
of target material. 
The Galactic density profile of ISM perpendicular to the Galactic plane at the relevant 
galactocentric distances of 1--3~kpc was constrained as a single Gaussian with a full width at half maximum 
of less than 200~pc and virtually no gas above 400~pc \citep{1984ApJ...283...90L}. 
At a distance of 5.5~kpc, \terzan\ would be at an offset above the disk of 160~pc and thus in an ambient gas 
density of a few times 0.1~cm$^{-3}$. 
The total energy in non-thermal electrons required for the emission of 2\ergs{33} of diffuse 
X-ray emission would be about 9$\times$10$^{49}$(n$_\mathrm{H}$/0.1~cm$^{-3}$)~erg if an electron energy of 
20~keV is assumed (Ok07). 
In contrast to the asymmetric morphology detected by Ok07 for GCs in a bow-shock scenario, 
we did not find evidence of a non-uniform shape of the excess emission from \terzan . 
This scenario could be tested by the presence of target material in the environment of the GC. 
Target material in the form of molecular clouds could be probed by carefully examining 
molecular emission lines that are shifted by the relative Galactic rotation velocity at the
physical location of \terzan .

\subsection{Thermal contribution}
The very high fitted temperature ($>$ 15 keV) of the diffuse X-ray
emission would suggest a non-thermal origin. However, at least a thermal
contribution to the total excess cannot be excluded at this point. 
If thermal bremsstrahlung is presumed as the emission mechanism, the
temperature of the plasma can be estimated from the X-ray luminosity, the
volume of the emission region, and the density of the plasma \citep{1995ApJ...451..200K}. 
With the radius of the emission region set to 5~pc (which corresponds to 188" at a distance of
5.5~kpc), this leads to 

$$\left(\frac{T}{10^7 \, \mathrm{K}} \right) \approx \left( \frac{L_{\rm X}^{\rm th}}{1.3\times 10^{33} (n/ 0.1 \, \mathrm{cm}^{-3})^2} \right)^2.$$

Assuming that the observed emission is entirely thermal, i.e., $L_{\rm X}^{\rm th}$ = $L_{\rm X}$, 
provides an upper bound for the temperature of the plasma.
It appears that this upper bound strongly depends on the plasma density. 
For a typical density at the GC position of 0.1~cm$^{-3}$ 
(see Sect. 3.4), the upper bound on the temperature is about $10^7$~K ($\approx$ 1~keV). 
Only for densities lower than 0.05~cm$^{-3}$ can the temperature exceed 15~keV. 

To heat plasma to such high temperatures, strong shocks would be
indispensable. 
The remnants of catastrophic events may release such strong
shocks (see, e.g., \citet{2007A&A...475..883A} for a remnant of a supernova Ia 
and \citet{2005A&A...444L..33D} for remnants of compact binary mergers). 
It was proposed that \terzan\ may host the required 
mergers; e.g., \citet{2002ApJ...571..830S} for white dwarf 
mergers and \citet{2006NatPh...2..116G} for neutron star -- neutron star mergers. 
However, even supernova remnants may have difficulty producing 
such high temperatures, because even the hot, thermal plasma in the young remnant of the
type Ia supernova remnant SN~1006 reaches a temperature of about 2~keV \citep[e.g.][]{2007A&A...475..883A}, 
significantly cooler than the temperature found for the thermal fit to the diffuse emission in \terzan . 

If the diffuse X-ray emission is indeed thermal, it could also originate in principle 
from a background galaxy cluster that by chance coincides with the core of \terzan . 
Since galaxy clusters with temperatures $>$10~keV are very rare \citep[e.g.][]{2002ApJ...567..716R}, 
such a correlation appears rather unlikely. 

From the available data, a contribution from thermal emission processes to the
measured flux cannot be ruled out, but it is not likely to represent the dominant fraction.

\section{Conclusions}
We discovered diffuse hard X-ray emission from the GC \terzan\ with a photon index of about 
1 and a peak flux density profile centered on the cluster core. 
The hard photon index makes a purely thermal emission scenario unlikely. 
Energetics would favor an SR scenario as the origin of the emission 
and would challenge simple IC and non-thermal Bremsstrahlung 
models generated by electrons accelerated by the bow shock of the GC. 
However, no simple model is clearly preferred to explain 
the observed emission, as expected from the limited statistics provided by the available
X-ray dataset. 
Additional X-ray observations, detailed multi-wavelength informations, 
as well as refined modeling are needed to accurately interpret the 
unique properties of the diffuse X-ray radiation in \terzan . 

\begin{acknowledgements}
This research has made use of data obtained from the Chandra Data Archive and software provided by the 
Chandra X-ray Center (CXC) in the application packages CIAO and ChIPS. 
We thank the referee for the very constructive feedback. 
\end{acknowledgements}

\bibliographystyle{aa}
\bibliography{}

\end{document}